\documentclass[10pt]{article}

\pdfoutput=1

\usepackage[a4paper]{geometry}
\usepackage{amssymb}
\usepackage{amsmath}

\usepackage[latin1]{inputenc}
\usepackage[T1]{fontenc}

\usepackage{graphicx} 

\usepackage[super,sort&compress,comma]{natbib} 

\usepackage[french,english]{babel}

\usepackage{setspace}

\begin{document}

\begin{center}

\Large{\bf Non-symmetric localized fold of a floating sheet} 
\vskip 2mm
\large{Marco Rivetti$\,^a$ \footnote{Present address: Surfaces du Verre et Interfaces, UMR 125, Saint-Gobain \& CNRS, 93303 Aubervilliers, France. email: \tt{marco.rivetti@saint-gobain.com} }}

\vskip 1mm
\normalsize{$^a$ Institut Jean le Rond d'Alembert, UMR 7190, \\
Universit\'e Pierre et Marie Curie \& Centre National de la Recherche Scientifique, \\
4 place Jussieu, 75005, Paris, France}


\end{center} 

\begin{abstract}
An elastic sheet lying on the surface of a liquid, if axially compressed, shows a transition from a smooth sinusoidal pattern to a well localized fold. This wrinkle-to-fold transition is a manifestation of a localized buckling. The symmetric and antisymmetric shapes of the fold have recently been described by Diamant \& Witten (2011), who found two exact solutions of the nonlinear equilibrium equations. In this Note, we show that these solutions can be generalized to a continuous family of solutions, which yields non symmetric shapes of the fold. We prove that non symmetric solutions also describe the shape of a soft strip withdrawn from a liquid bath, a physical problem that allows to easily observe portions of non symmetric profiles.

\end{abstract}

\section{Introduction}
\label{intro}

Localization of stress or strain is a phenomenon that appears in wide range of mechanical systems, like for example  granular media \cite{anand2000}, fluid-saturated porous media \cite{loret1991}, twisted elastic rods \cite{nizette1999} or damage of materials \cite{bazant1988}, \cite{pham2011}. 
As extreme consequence, localization leads to the formation of singularities, like fractures \cite{aifantis1992}, break-up of liquid threads \cite{eggers1997} or d-cones in crumpled papers \cite{ben-amar1997}, just to cite a few examples. 

In the field of elastic structures, the possible localization of the deformation of an axially compressed beam resting on an elastic foundation is a problem which has attracted a large interest. 
%
For this kind of systems, linear post-buckling r\'egime is characterized by a sinusoidal shape with a well defined wavelength $\lambda$, provided that the length of the beam is much larger that $\lambda$ \cite{timoshenko_strength_mat}.  However, it has be shown that the nonlinear post-buckling response exhibits a complex behavior, depending on the constitutive law of the foundation and thus on the type of bifurcation that occurs \cite{hunt1989}, \cite{champneys1997}. For instance, thin films on a soft elastic substrate display cascades of space-period doubling eventually leading to chaos \cite{brau2011}, as well as localized folding, restabilisation, delamination or creasing of the film \cite{ebata2012}, \cite{li2012}, \cite{wadee2002}.

In the limit case of a thin sheet lying on a liquid surface, it has be shown that the sinusoidal wrinkles appearing close to the buckling threshold (i.e for low confined configurations) spontaneously transform into a localized fold for highly confined configurations \cite{pocivavsek2008}. This phenomenon  covers a wide range of length scales, spanning from lipid monolayers \cite{baoukina2008} up to ice pack in the ocean \cite{sjolind1985}, and may lead to the delamination of the film from the liquid \cite{wagner2011}. 
Audoly \cite{audoly2011} has derived an amplitude equation for the envelope of the pattern, explaining the localization of the buckling and relating it to the one on a nonlinear elastic foundation. 
In a recent paper, Diamant \& Witten \cite{diamant2011} have shown that the nonlinear equation governing the shape of the fold is integrable in the case of an infinite long sheet, and have proposed two exact solutions describing symmetric and antisymmetric shapes. 

In this Note, we show that symmetric and antisymmetric solutions are two particular cases included in a continuous family of solutions of the nonlinear equation. This family of solutions describes, in general, non symmetric shapes of the fold, all sharing the same total energy. 
While for the buckled floating sheet all the solutions are possible, we show that for a slight different system, consisting in the extraction of an elastic strip from a liquid bath, the exact shape of the strip is described by just one given solution among the family.  





\section{A continuous family of solutions}

\begin{figure}[h]
\begin{center}
\setlength{\unitlength}{0.9cm}
\begin{picture}(13,4.1)
	\put(0,0.1){\includegraphics[width=11.7cm]{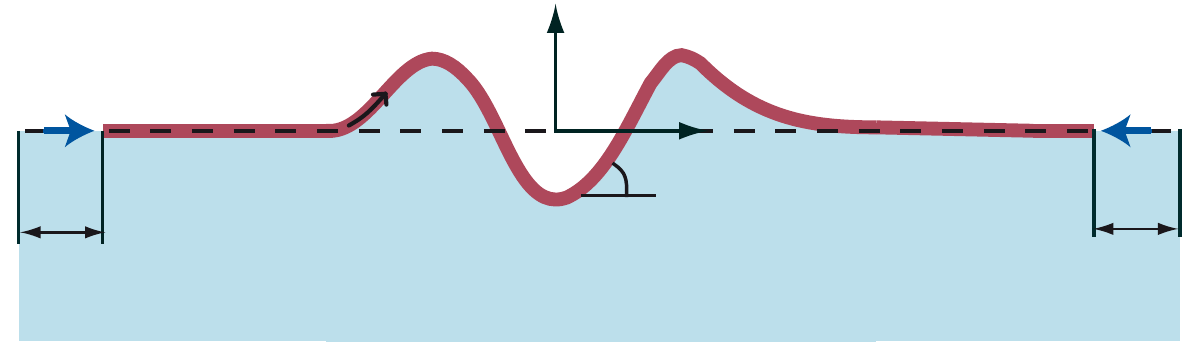}}
	\put(7.4,2.6){\large{$x$}}
	\put(6.3,3.5){\large{$y$}}
	\put(3.8,2.8){\large{$s$}}
	\put(7.,1.85){\large{$\theta$}}
	\put(0.5,1.5){\large{$\Delta$}}
	\put(12.3,1.5){\large{$\Delta$}}
	\put(0.4,2.6){\large{$P$}}
	\put(12.4,2.6){\large{$P$}}
\end{picture}
\caption{2D schematic view of a fold in a buckled floating beam. The angle $\theta$ (between the tangent to the beam and the horizontal) as well as the horizontal $x$ and vertical $y$ positions, are expressed as a function of the arc-length $s$. The compressive dead load is noted $P$.}
\label{fig:notations}
\end{center}
\end{figure}
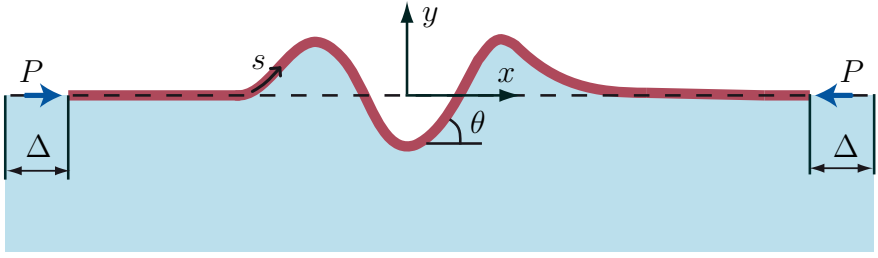

Consider the problem of a flexible and inextensible sheet of length $2L$, width $w$ and thickness $e$ (with $L\gg w \gg e$), lying at the surface of a liquid (Fig. \ref{fig:notations}). The bending rigidity of the sheet is $EI$, $E$ being Young modulus and $I = we^3/12$ the second moment of the cross area of the sheet. The density of the liquid is $\rho$, and $g$ is the acceleration of gravity. 
We consider that the deformations of the sheet are confined in the $(x,y)$ plane, which leads to study the equilibrium of a 2D elastic beam. 
The beam is parametrized by its arc-length $s \in (-L,L)$.  Horizontal $x$ and vertical $y$ positions of an infinitesimal element $\mathrm{d}s$ can be related to the angle $\theta$ of the element with respect to the horizontal by: 
\begin{align}
x'(s) &=  \cos \theta (s) \label{eq:xprime} \\
y'(s) &=  \sin \theta (s) \label{eq:yprime} \, ,
\end{align}
where prime $'=\mathrm{d}/\mathrm{d}s$ denotes the derivative with respect to the arc-length.

The linear response of this system to an axial compressive load $P$ is a buckled sinusoidal configuration, with a well defined wavelength $\lambda \sim \left( \frac{EI}{\rho g} \right)^{1/4} $, provided that $L \gg \lambda$ \cite{timoshenko_strength_mat}. The linear response is valid only for a low confined beam (i.e. for small values of the displacement $\Delta$ at the ends of the beam).  

We introduce $\lambda$ as the reference length-scale, which is equivalent to take $EI= \rho g =1$. We thus consider that all variables in the following are dimensionless.
Here, as we are interested in the transition from a wrinkle state to a localized fold, we let $L \to \infty$ and we make the assumption that all deformations vanish at infinity ($y=\theta=\theta' =0$ at infinity). This leads to study the homoclinic orbit of the system, with the ends of the beam which represent the homoclinic connection in the phase space $(\theta,\theta')$ \cite{champneys1998}.  
For these reasons, the fold appearing in the middle of the beam (which is flat otherwise) can be seen as a boundary layer solution. 
%

%
%
For an exhaustive derivation of the leading equations and a description of the nonlinear response, the reader can refer to the papers of Audoly \cite{audoly2011} or Diamant \& Witten \cite{diamant2011}.
The total energy of the system, which is the sum of elastic, potential and compressive energy, writes:
\begin{equation}
\mathcal{E}(\theta, \theta',y,y') = \int_{-\infty}^\infty \left( \frac{1}{2} \theta' (s) ^2 + \frac{1}{2}  y(s)^2 \cos \theta (s) - P (1-\cos \theta (s)) \right) \mathrm{d}s \;.
\label{eq:energy}
\end{equation}
%

Functions $\theta(s)$ and $y(s)$ are related by equation (\ref{eq:yprime}), and the correct way to look for the equilibrium equations is to find the stationary points of the lagrangian $ \mathcal{L} = \mathcal{E} - \int q(s) (y'(s) - \sin \theta (s) ) \mathrm{d}s $, with $q(s)$ a local lagrangian multiplier. This leads to the system of nonlinear equilibrium equations:
\begin{equation}
\left\{
\begin{aligned}
&\theta '''(s) + \frac{1}{2} \theta '(s) ^3 + P \theta '(s) + y(s) = 0 \\
&y'(s) = \sin \theta (s) 
\end{aligned}
\right.
\label{eq:system_fold}
\end{equation}  

It is possible, by deriving the first equation in (\ref{eq:system_fold}) once, to write one differential equation for $\theta(s)$:
\begin{equation}
\theta''''(s) + \frac{3}{2}\theta'(s)^2 \theta''(s) + P \theta ''(s) + \sin \theta (s) = 0 \,.
\label{eq:theta4_fold}
\end{equation}

Despite its complexity, Diamant \& Witten \cite{diamant2011} were able to find two exact solutions to equation (\ref{eq:theta4_fold}) which satisfy the boundary conditions for  $\theta$ vanishing  at infinity. These solutions  are: 
\begin{align}
\theta_s (s) = -4 \arctan \left[ \frac{c \sin (k \, s)}{k \cosh (c \, s) }\right] \\
\theta_a (s) = -4 \arctan \left[ \frac{c \cos (k \, s)}{k \cosh (c \, s) }\right]  
\end{align}
with coefficients $c$ and $k$ linked to the load $P$ by $k = \sqrt{2+P} /2$ and $c= \sqrt{2-P}/2$. Note that also $-\theta_s$ and $-\theta_a$ are solutions. 
Here, $\theta_s$ is the solution which gives a symmetric shape and $\theta_a$ an antisymmetric shape of the fold. 
Indeed, $\theta_s$ is such that it vanishes at $s=0$ as well as all its even derivatives $\theta_s''$, $\theta_s''''$, ... In contrast, $\theta_a$ is such that all the odd derivatives $\theta'_a$, $\theta'''_a$, ... vanish at $s=0$.

However, these solutions are not unique, and it is possible to verify, by simple substitution into equation (\ref{eq:theta4_fold}), that all the functions of the form:
\begin{equation}
\theta_\phi (s) = -4 \arctan  \left[ \frac{c \sin (k  (s+\phi))}{k \cosh (c \, s) }\right] 	
\label{eq:theta_phi}	
\end{equation}
are equilibrium solutions of the problem (satisfying also the boundary conditions).
The parameter $\phi$ is a real number, and it appears only in the sine, so that it is not a mere change of variable for $s$. 
The fact that $\phi$ can take any real values yields that equation (\ref{eq:theta_phi}) is a continuous family of solutions. 
Hence, $\theta_s$ and $\theta_a$ are two particular solutions belonging to this family, recovered when $\phi = 0$ and $\phi = \pi/2k$, respectively. 
For $\phi \neq n\pi / 2k$  ($n$ being an integer), the solution $\theta_\phi$ does not have any kind of symmetry, as we will show explicitly in the next paragraph. 



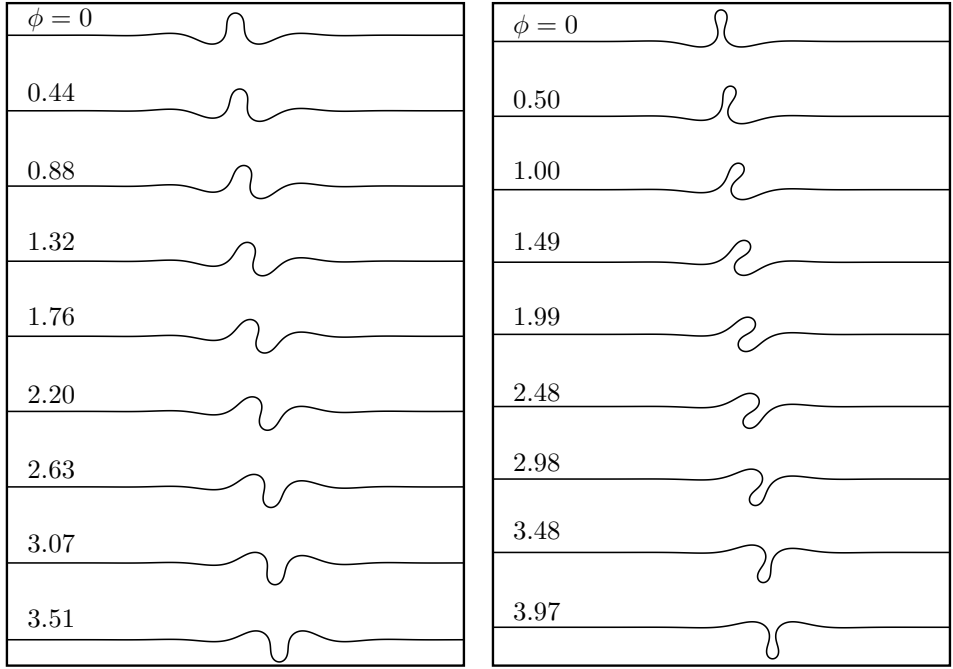
\begin{figure}[h]
\begin{center}
\setlength{\unitlength}{0.9cm}
\begin{picture}(14,10.1)
	\put(0,0){\includegraphics[width=6.19cm]{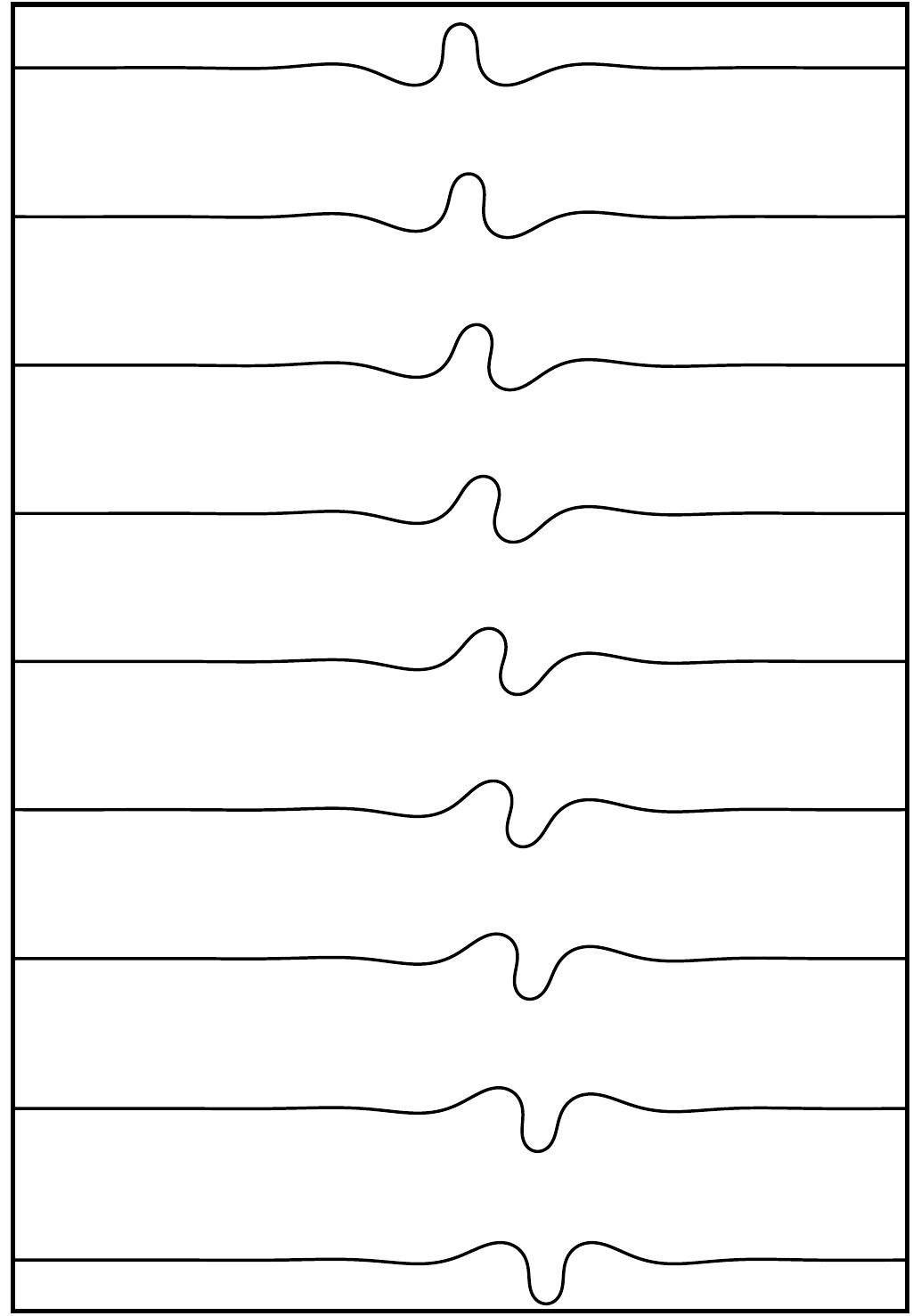}}
	\put(7.1,0){\includegraphics[width=6.19cm]{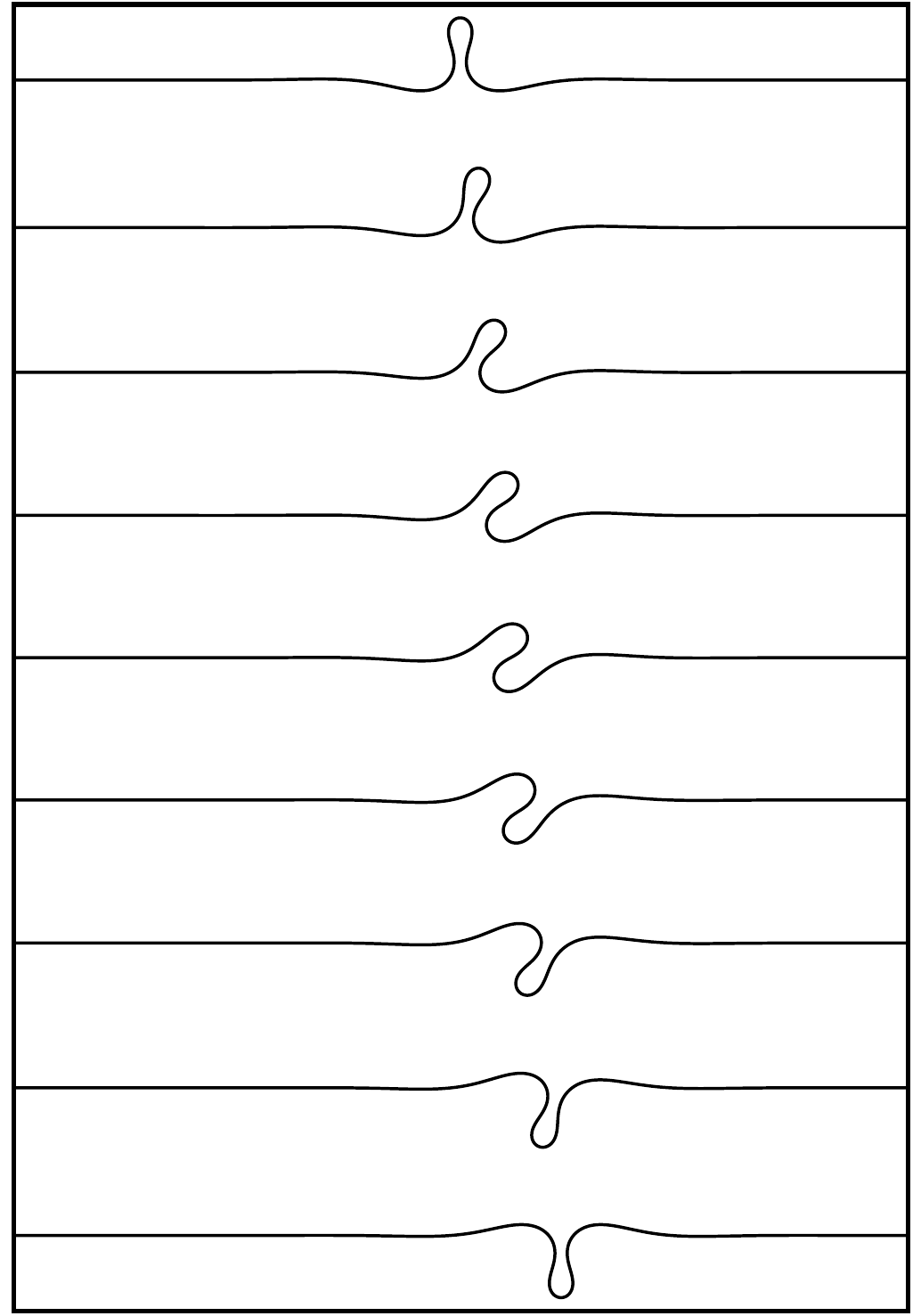}}
	\put(0.4,9.45){$\phi = 0$}
	\put(0.4,8.35){$0.44$}
	\put(0.4,7.2){$0.88$}
	\put(0.4,6.15){$1.32$}
	\put(0.4,5.05){$1.76$}
	\put(0.4,3.9){$2.20$}
	\put(0.4,2.8){$2.63$}
	\put(0.4,1.7){$3.07$}
	\put(0.4,0.6){$3.51$}
	\put(7.5,9.35){$\phi = 0$}
	\put(7.5,8.25){$0.50$}
	\put(7.5,7.2){$1.00$}
	\put(7.5,6.15){$1.49$}
	\put(7.5,5.05){$1.99$}
	\put(7.5,3.93){$2.48$}
	\put(7.5,2.9){$2.98$}
	\put(7.5,1.9){$3.48$}
	\put(7.5,0.7){$3.97$}
\end{picture}
\caption{Nine examples of solutions $(x_\phi,y_\phi)$ included into the family of solutions given by equation (\ref{eq:theta_phi}). The solutions correspond to $P=1.2$ (left) and $P=0.5$ (right). The corresponding value of $\phi$ is given for each profile.}
\label{fig:famille_solutions}
\end{center}
\end{figure}

The shape $(x_\phi,y_\phi)$ of the fold can be obtained by integration of equations (\ref{eq:xprime})-(\ref{eq:yprime}). Figure \ref{fig:famille_solutions} shows, for two different loads $P_1=1.2$ (left) and $P_2=0.5$ (right), nine shapes belonging to the family of solutions. The corresponding values of $\phi$ for each profile are given in the figure. By observing the shapes, one can easily remark that, beside the symmetric and antisymmetric profiles, there exists infinite non symmetric profiles. 

Diamant \& Witten \cite{diamant2011} already pointed out that symmetric and antisymmetric solutions have the same energy $\mathcal{E}$. By substituting $\theta_\phi$ and $y_\phi$ into equation (\ref{eq:energy}), it results that $\mathcal{E} = \frac{8}{3} \sqrt{2-P}(2-P)$ : the functional no more depends on $\phi$ and all the profiles have the same energy. 

\section{Selecting one non symmetric solution}

\begin{figure}[h!]
 \begin{center}
 \setlength{\unitlength}{0.9cm}
 \begin{picture}(13,3.9)
 	\put(0,0){\includegraphics[width=11.7cm]{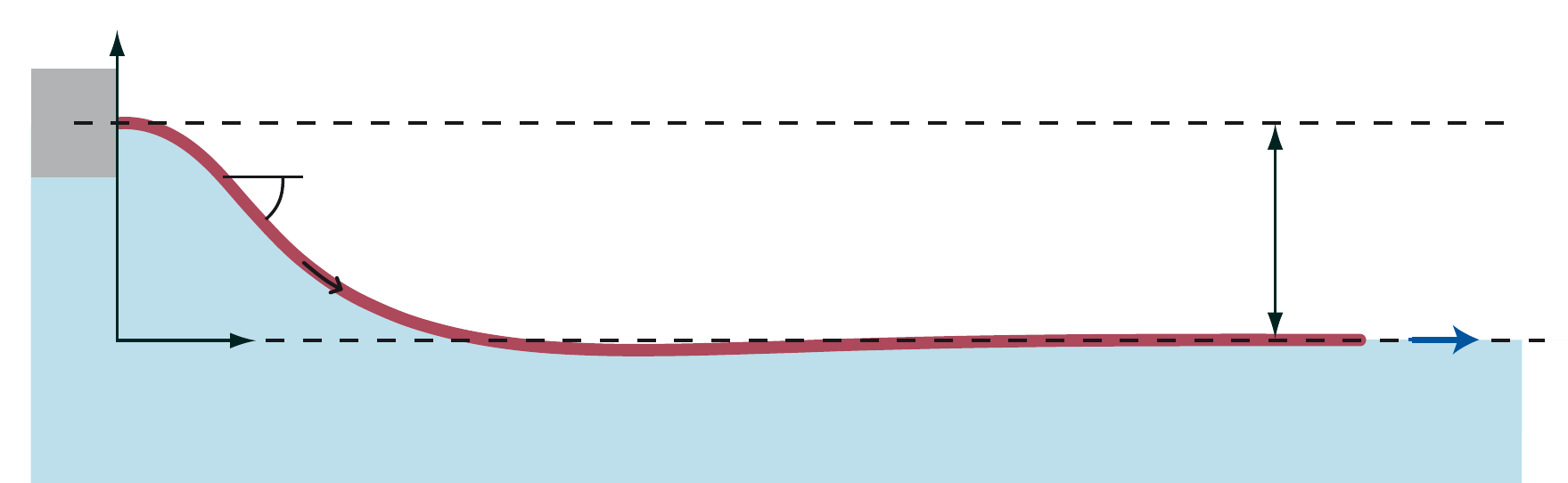}}
	\put(2,1.3){\large{$x$}}
	\put(1.1,3.5){\large{$y$}}
	\put(10.7,2){\large{$h$}}
	\put(11.8,1.4){\large{$\gamma$}}
	\put(2.4,2.1){$\theta$}
	\put(2.8,1.8){\large{$s$}}
 \end{picture}
 \end{center}
 \caption{Schematic view of the extraction of a 2D elastic strip from a liquid bath. The strip is clamped at one end. The distance between the highest point of the strip and level of the liquid is noted $h$, and it is the control parameter of the extraction.}
 \label{fig:schema_menisque}
 \end{figure}
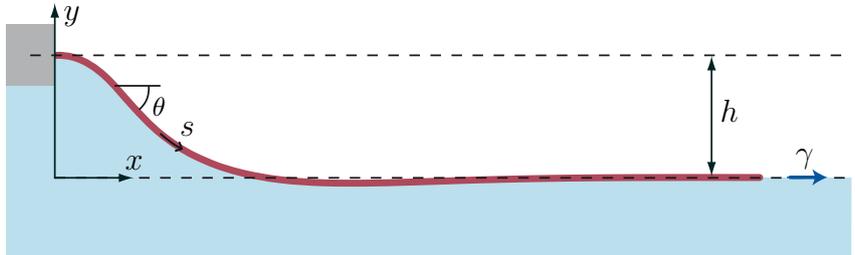

Among all the configurations parametrized by  $\phi$, the question to know which ones are stable (and therefore observable) is still open. The fact that all the solutions have the same energy implies that they are neutrally stable with respect to each other. Nevertheless, in order to conclude about the the stability of an equilibrium solution, the sign of the second variation of the energy should be tested with respect to \textit{any} possible infinitesimal path around \textit{any} equilibrium solution (this sign being zero along the path which connects all the solution). 
The experimental profiles shown by Pocivavsek \& al. \cite{pocivavsek2008} in some cases do not look completely symmetric, suggesting the fact that some non symmetric solutions may be stable, but the lack of investigations on this topic does not allow for an exhaustive answer. 

It exists a slightly different physical problem, in which the shape of the beam is exactly described by one given function $\theta_\phi$, the right value of $\phi$ depending on a control parameter.
In Rivetti \& Antkowiak \cite{rivetti_meniscus}, the authors study the equilibrium configuration of a 2D inextensible strip extracted from a liquid bath (Fig. \ref{fig:schema_menisque}). In the experimental setup, executed in a rigid box with glass walls, a narrow elastic strip is clamped at one end and lies at the free interface between the liquid and the air. The liquid is quasi-statically withdrawn from the box, so that the control parameter in the problem is the height of extraction, $h$, defined as the distance between the clamp and the (decreasing) level of the liquid interface. The parameter $h$ can be tuned starting from zero (a completely flat strip) up to a maximal value at the moment of extraction. In this system, elasticity and hydrostatic pressure interact as for the buckled floating beam. However, in this problem there is no axial compressive load, but a pulling force at the end of the strip because of surface tension $\gamma$.

In the case in which the length of the strip is $L \gg \lambda$, the deformation localizes in a boundary layer (near the clamp) and is nil elsewhere, as in the buckling problem described in the previous section. The shape of the strip in the boundary layer is obtained solving the system of equilibrium equations: 
\begin{equation} 
\left\{
\begin{aligned}
	&  \theta '''(s) + \frac{1}{2}{\theta'}^3  - \zeta^2 \theta'(s)    +  y(s)  = 0 \\
	& y'(s) = \sin \theta (s)  
\end{aligned} 
\right.
\label{eq:system_meniscus}
\end{equation}
with $\zeta^2 = L_\mathrm{gc}^2 / \lambda^2$ representing the contribution of surface tension ($L_\mathrm{gc} = \sqrt{ \gamma / \rho g} $ is the capillary length) \cite{rivetti_meniscus}. The minus sign for $\zeta^2$, with respect to the compressive load $P$ in equation (\ref{eq:system_fold}), is justified by the fact that surface tension pulls the strip. 

\begin{figure}
 \begin{center}
 	\includegraphics[width=0.96\textwidth]{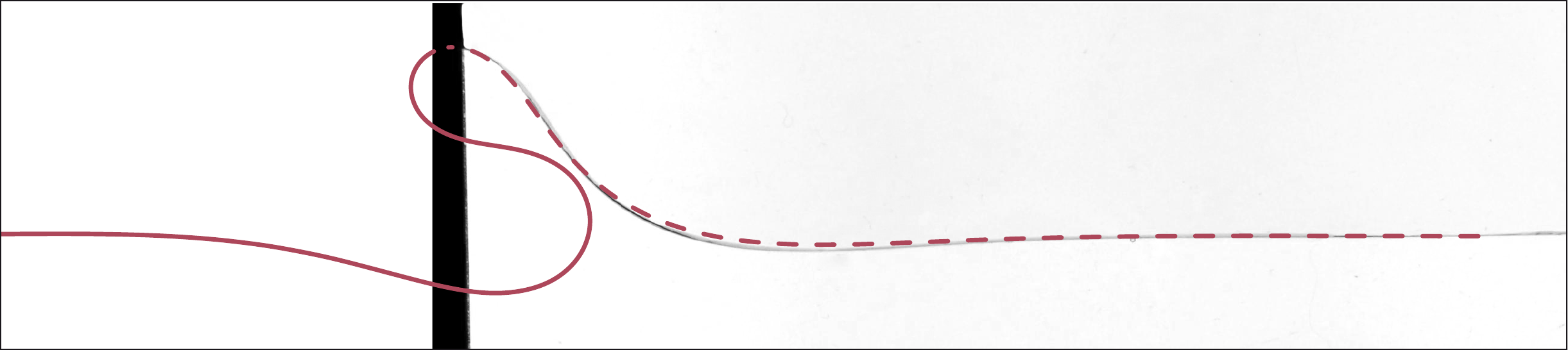}
 \end{center}
 \caption{Comparison  between the experimental and the theoretical shape of an elastic strip withdrawn from a liquid bath. Here, the ($x,y$) theoretical profile (red curve) is obtained for $h=1.91 \lambda$ and $\zeta=0.24$, leading to $P=-0.06$ and $\phi = 1.33$. The physical strip corresponds to a portion (the red dashed curve) of a non symmetric fold (the other portion of the fold is shown with a red continuos line). }
 \label{fig:compar_menisque}
 \end{figure}

The system of equations (\ref{eq:system_meniscus}) can be verified by all the solutions $\theta_\phi$ given by equation (\ref{eq:theta_phi}), if the substitution $P=-\zeta^2$ is done. 
Boundary conditions at $s \to \infty$, reconnecting the localized deformation to a flat external solution, are identical to those of system (\ref{eq:system_fold}) and thus satisfied by $\theta_\phi$. However, the presence of a clamp requires to pay attention on the boundary conditions at $s=0$. The condition of zero angle at the clamp, $\theta(0)=0$, can be settled by a simple change of variable $s \to s - \phi$ in equation (\ref{eq:theta_phi}), which transposes $\phi$ into the hyperbolic cosine: 
\begin{equation}
\bar{\theta}_\phi = -4 \arctan  \left[ \frac{c \sin (k  s)}{k \cosh (c \, (s - \phi)) }\right] 	\;.
\end{equation}
Nevertheless, the control parameter explicitly appears in the last condition, $y(0) =h$.
Satisfying this condition leads to a selection of one specific value of $\phi$: 
\begin{align}
y(0) = -\int_0^\infty &\sin \bar{\theta}_\phi (s) \, \mathrm{d}s =2\sqrt{2-P} \; \mathrm{sech} \left( \frac{1}{2} \phi \sqrt{2-P} \right) = h 	\label{eq:h_vs_phi} \\
& \Leftrightarrow \; \phi =\frac{2}{\sqrt{2-P}} \, \mathrm{sech}^{-1} \left( \frac{h}{2 \sqrt{2-P}} \right)		\label{eq:phi_vs_h}
\end{align} 
%
Hence, for any value of $h$, the shape of the elastic strip is given by one given solution $\bar{\theta}_\phi$, which is a portion of a (generally) non symmetric fold. The Fig. \ref{fig:compar_menisque} shows good agreement between the experimental profile and the corresponding solution obtained for $P = -\zeta^2 = -0.06$ and $\phi = 1.33$. 

From the analytical solution, it is therefore possible to obtain the value of the physical parameters characterizing the extraction, as for instance the value of the curvature at the clamp, $\theta'(0) = -h$, or the vertical force needed to withdrawn the strip, $F_y = \int_0^\infty y(s) \cos \theta(s) \, \mathrm{d}s = \sqrt{(2+\zeta^2)h^2 - \frac{1}{4}h^4}$.

\section{Conclusion}

In this Note we considered the problem of a floating beam buckling under the action of a compressive load. 
In order to investigate the shape of the fold that spontaneously appears for high confined configurations, we presented a family of analytical solutions parametrized by a real parameter $\phi$. These new solutions predict non symmetric profiles of the fold, and they are a generalization of the symmetric and antisymmetric solutions already introduced  \cite{diamant2011}. We showed that all the profiles given by the solutions are neutrally stable with respect to each other, because they share the same total energy. 
However, the lack of an exhaustive study of the stability for all the  solutions with respect to any perturbation does not allow to answer the question of which shape is selected when the fold forms.
%

We proved that portions of non symmetric configurations can nevertheless be observed in a slightly different problem. Indeed, we related the problem of the buckling on a liquid surface to the problem of the extraction of a floating beam \cite{rivetti_meniscus}. Although the equilibrium equations for both systems are very similar, boundary conditions in the latter are responsible for the selection of only one given solution. All the non symmetric profiles can therefore be observed by tuning the value of a control parameter appearing in the boundary conditions. 
%

One may ask whether the generalization introduced in this paper for the floating beam could apply to other soliton solutions, as in Sine-Gordon or Korteweg-de Vries equation. It is straightforward to prove for both these cases that a solution keeps its validity if a $\phi$-shifting is applied like in this Note; however, as Sine-Gordon and Korteweg-de Vries equations are PDE, the $\phi$-shifting merely leads to a change of variable for space or time, without perturbing the symmetry of the solution. Indeed, the intrinsic interest of the $\phi$-shifting in the floating beam is that the equilibrium equation is an ODE, and $\phi$-shifting effectively affects the symmetry of the solution.  

This Note has proved that symmetric and antisymmetric solutions were not general enough. However, more general solutions than those introduced here could exist, because a direct resolution of the equilibrium equation has not be done yet. 
%
Moreover, a step forward in the comprehension of the problem may be the extension of the homoclinic solution to other orbits in the phase diagram, for instance to determine the exact shape of a buckled floating beam with a finite length.

\section*{Acknowledgements}

The author is grateful to the suggestions and the encouragements of Arnaud Antkowiak and S\'ebastien Neukirch during the preparation of this article. 





\bibliographystyle{crasstyle}
\bibliography{biblio_cras}

\end{document}